# D-Mag – a laboratory for studying plasma physics and diagnostics in strong magnetic fields

**B. Jagielski,**[a,1] **U. Wenzel,**[a] **T. Sunn Pedersen,**[d] **A. Melzer,**[b] **A. Pandey,**[a] **F. Mackel**[c] **and the Wendelstein 7-X team**

[a] *Max-Planck-Institute for Plasma Physics (IPP), EURATOM Association,*
*Wendelsteinstr. 1, 17491 Greifswald, Germany*

[b] *University Greifswald, Institute for Physics,*
*Felix-Hausdorff-Straße 6, 17489 Greifswald, Germany*

[c] *Max-Planck-Institute for Plasma Physics (IPP),*
*Boltzmannstraße 2, 85747 Garching bei München, Germany*

[d] *Type One Energy Group, Madison, WI USA 53703*

*E-mail*: bartholomaeus.jagielski@ipp.mpg.de

ABSTRACT: We have set up a diagnostic magnet (D-Mag) laboratory for a wide range of applications in plasma physics. It consists of a superconducting magnet for field strengths of up to 5.9 T. The main purpose is to provide an experimental environment for the development of plasma diagnostics for nuclear fusion studies and the investigation of dusty plasmas in strong magnetic fields. We describe in the article the setup and operation of the D-Mag. Some applications are presented for the development of plasma diagnostics, such as neutral pressure gauges and Langmuir probes that have to be operated in strong magnetic fields. Among the examples is the test of the long-pulse capability and stability of the diagnostic pressure gauge (DPG) for the ITER device.

---

[1] Corresponding author.

# Contents

## 1. Introduction

The investigation of various materials and diagnostics in strong magnetic fields plays an important role in many different scientific and industrial fields. For example, magnetization effects of dusty plasmas [1-3] are relevant to astrophysics, as well as accurate temperature measurements using thermocouples [4], where a magnetic field can induce voltage in the wire and cause inductive heating, are of great significance. In particular tests in strong magnetic fields are necessary to develop and validate diagnostics for current and future experimental fusion devices, such as Wendelstein 7-X (W7-X), ITER and DEMO. All of the diagnostics installed in the plasma vessel will be exposed to magnetic fields of up to 2.5 T (W7-X) [5] and 8 T (ITER) [6] for detectors at the divertor, must withstand the strong Lorentz force and require a specific characterization in respect to the magnetic field. For example, the field strength at the location of the pressure gauges in W7-X ranges from 1.64 to 2.19 T [7]. Especially the targeted long pulse and continuous mode operation of the fusion devices places high requirements on the detectors. In W7-X for instance, the translational mechanism, driving Langmuir probes towards the plasma edge and retracting them, is subjected to high thermomechanical loads in the magnetic field and pressure gauges, such as the newly developed crystal cathode pressure gauges (CCPG) [8], must maintain their operation under the influence of the Lorentz force. In order to qualify a robust operation, the instruments are exposed in versatile and accessible test facilities and the D-Mag facility is set up for this purpose.

The central component of the D-Mag laboratory is the diagnostic superconducting magnet (D-Mag), which is operated in a non-persistent mode, allowing the magnetic field strength to be adjusted. Originally used as a 140 GHz gyrotron, the cryomagnet was given its second life in 2016 at the Physics Institute of the University of Greifswald, Germany. After dedicated modifications and extensions of the ultra-high vacuum (UHV) apparatus and the diagnostics of the setup, the D-Mag laboratory provides an excellent opportunity to analyze different detectors within variable magnetic field strengths of up to 5.9 T and in a wide pressure range from the $10^{-9}$ mbar UHV-region to atmospheric pressure.

## 2. Experimental setup

### 2.1 Diagnostic magnet (D-Mag)

Figure 1 and figure 2 show the experimental setup of the D-Mag laboratory. It consists of several elements: a superconducting magnet, the analysis chamber, gas-inlet system, pump system (turbomolecular pump combined with a scrollpump) and the diagnostic systems (residual gas analyzer (RGA), pressure gauges, infra red (IR) camera and pyrometer). All connecting pipes, UHV cross chambers and the analysis chamber are covered with several heating mats, which are wrapped with aluminum foil. This allows targeted heating of individual segments of the apparatus, controlled by two bake out control units (Prevac BCU 14).

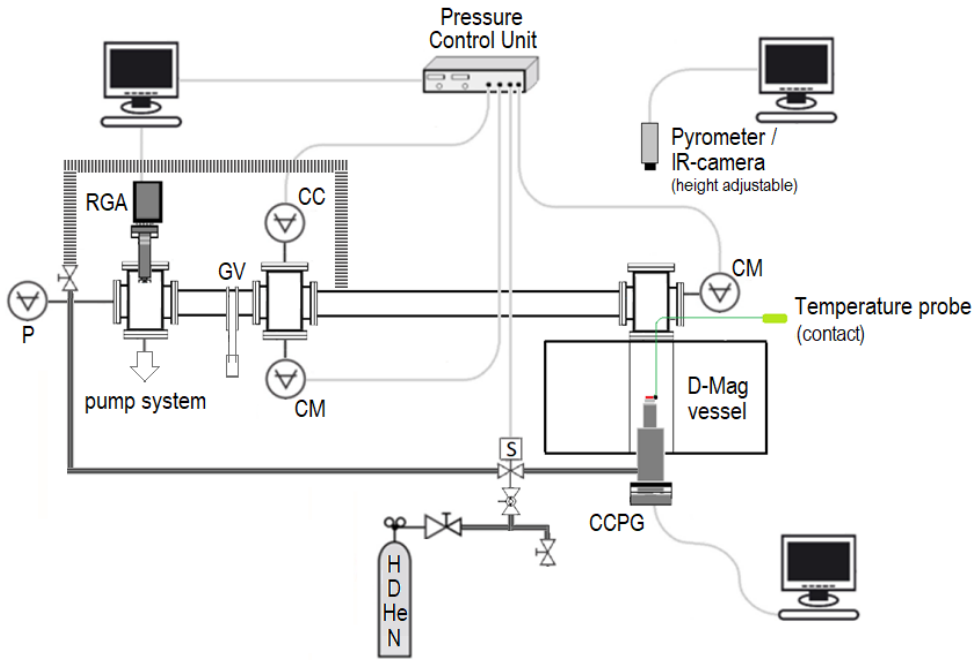

**Figure 1.** Schematic setup of the D-Mag laboratory. An analysis chamber is placed inside a bore of the D-Mag vessel (figure 3). Here: a pressure gauge (CCPG) is inserted into the analysis chamber. The pumping stand, pressure gauges and residual gas analyzer are installed in some distance to the magnet in order to avoid a negative influence of a magnetic stray field. Both capacitance manometers/Baratrons® (CMs) and the cold cathode (CC) must be attached to the central UHV double cross chamber when the magnet is operated. A pneumatic UHV gate valve (GV) is placed between the pumping stand and the double cross chamber as a manipulator for the gas conductance: while the GV is closed, gas is pumped out through the corrugated hose. A residual gas analyzer (RGA) is installed at the cross chamber above the pumping stand.

We describe in the following subsection the setup of the D-Mag laboratory and its diagnostics.

### 2.1.1 Specifications of the D-Mag

The superconducting magnet was manufactured by OXFORD Instruments and was part of a 140 GHz gyrotron. The gyrotron tube was replaced with a custom-built and heatable analysis chamber for the characterization of instruments such as pressure gauges in different gases (figure 3).

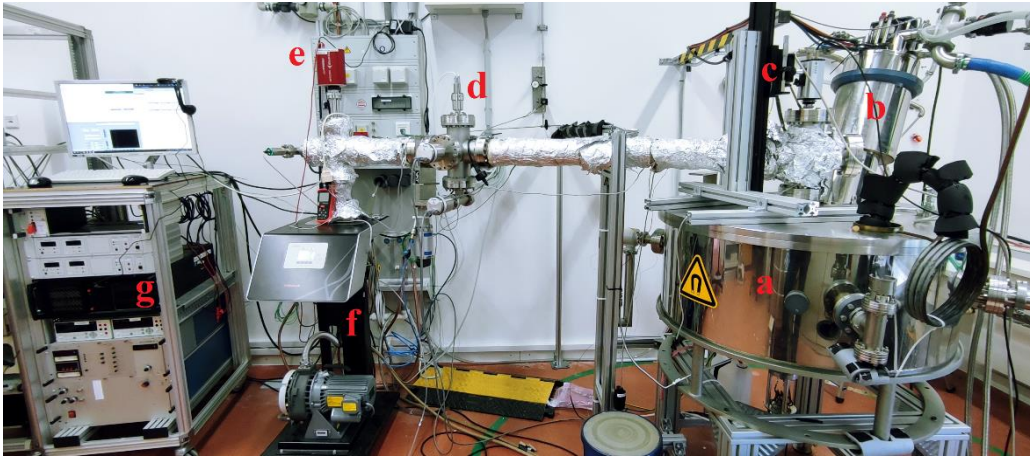

**Figure 2.** Photograph of the D-Mag laboratory: a) D-Mag vessel, b) cryogenic coldhead, c) camera-stand (here: DIAS pyrometer), d) double cross-chamber with two Baratrons® and a cold cathode gauge, e) RGA, f) pump station, g) rack with electronic equipment for the control of the magnet.

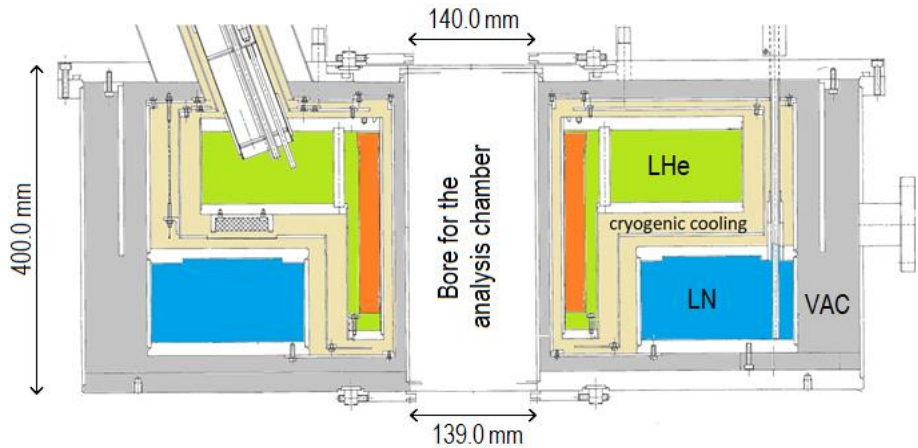

**Figure 3.** Scheme of the axisymmetrical cut of the D-Mag vessel. The analysis chamber is inserted into the bore in the center. The reservoirs for liquid helium (LHe), liquid nitrogen (LN) and the insulation vacuum (VAC) are shown. The solenoid is shown in orange.

The solenoid is 20 cm high, with 10500 turns, and made of a multifilamentary, copper stabilized niobium titanium (NbTi) super-conducting coil, with a transition temperature of 10.6 K. The magnet can only be operated in non-persistent mode, i.e. electrical current must be supplied continuously to maintain the magnetic field. The operation mode allows free adjustment of the magnetic field strength (B) up to a maximum of $B_{max}$ = 5.9 T in the center of the solenoid, at 92 A coil-current and an operating temperature of 4.2 K. With a nominal induction of 48 H, this results in an energy for the maximal magnetic field of E = 203 kJ. The magnetic stray field in the vertical plane as a function of the solenoid current is shown in figure 4 and figure 5. The manufacturer's specifications regarding the magnetic field distribution suggest that in the field in the magnetic field center of the magnet is almost homogeneously distributed in a small region ($\pm$ 3 cm). B drops rapidly further away from the center. At a vertical distance of ~140 cm from the top lid of the vessel, B falls to < 4 mT, so that additional detectors, such as camera systems, can be safely mounted at this height. For specific applications reported below, a pyrometer and an IR camera can still be operated at B $\leq$ 40 mT, which is achieved at a distance of 73 cm from the central magnetic field, or 53 cm above the D-Mag lid.

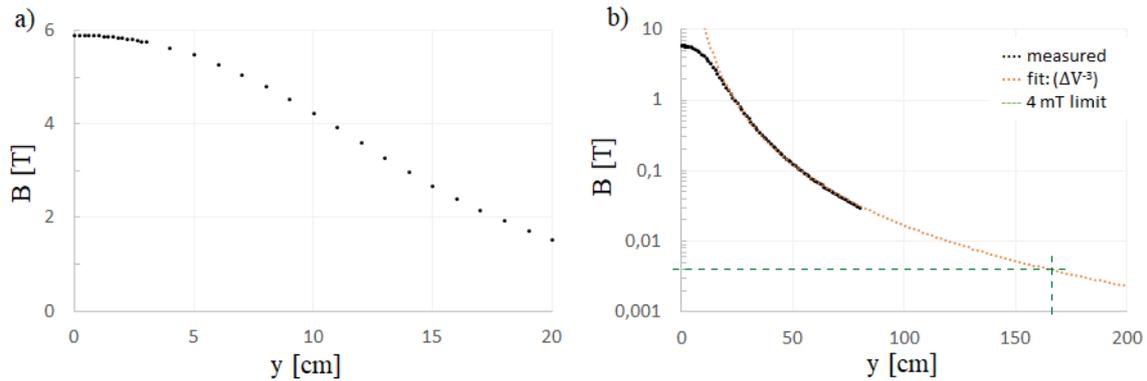

**Figure 4.** Magnetic field strength (B) as a function of the vertical distance (y) to the center of the magnet. a) The homogenous range of the magnetic field in the center is $\pm$ 5 cm (5.9 T to 5.5 T). b) At a vertical distance of 163 cm from the center of the magnet, the magnetic field drops to 4 mT, which is the suggested limit for operation of the installed pumps, electric pressure valves, Baratrons® and other electrical devices.

Before installing the analysis chamber, B was measured versus the coil current. In the operating instructions [10], the manufacturer specifies the coil current at $B_{max}$ and assumes a linear field-current ratio of 0.06367 T/A. However, a discrepancy was found between our values and the manufacturer's specifications that is shown in figure 5. Although the initial and final values were in good agreement, a discrepancy of up to 25% was measured in the range between 30 A and 60 A. This effect could be explained by an increased permeability and magnetization of the steel of the D-Mag vessel due to welding spots. Non-linear magnetization effects cannot be excluded, because the material characteristics of the D-mag body and its inner components are not known. Because of the relative permeability of the steel used for the analysis chamber ($\mu_r$ < 1,01), a magnetization of the chamber is not expected.

The solenoid is placed in a 26 l liquid helium (LHe) cryostat chamber, which is fixed axially and radially with suspension elements made out of glass fiber spacers. It is thermally insulated by a vacuum case of 110 l. The cryostat is precooled by a liquid nitrogen (LN) reservoir comprising 33 l LN. It has two radiation shields made of multilayered superinsulation of stainless steel and an aluminum alloy. The outer radiation shield is cooled by the LN reservoir, the inner one by a pulse tube cryocooler (0.8 NM torque), which is attached to the neck of the cryostat vessel, operated and cooled by a compressor (Leybold ARW 6000) with 16 bar helium pressure.

A nitrogen level probe is inserted into a port at the top plate of the vessel, which extends to the bottom of the LN-reservoir. Another open port serves as a nitrogen boil-off exhaust.

Electrical access to the magnet is provided on the cryostat service neck, where two LHe level probes and a carbon resistance temperature sensor (Allen-Bradley resistor) are mounted. The latter is monitored by a digital multimeter (Agilent 34410A 6 ½) and the LHe and LN levels are read out by two level meters (OXFORD ILM 211).

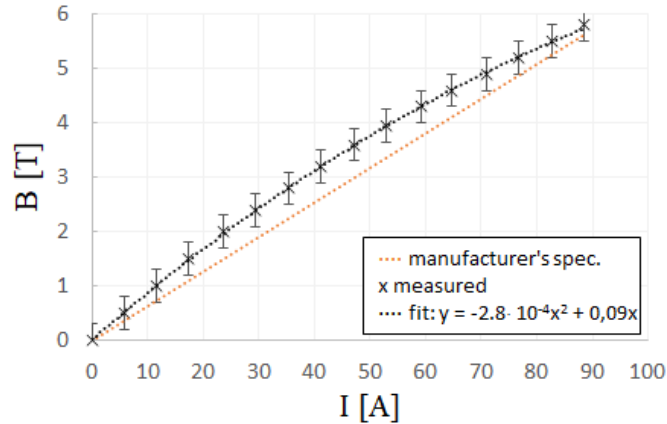

**Figure 5.** Magnetic field (B) at the center of the solenoid as a function of the solenoid current (I). The measured flux deviates from the calculated linear dependence of the manufacturer's specification, possibly due to the magnetization of the austenitic stainless steel walls of the D-Mag vessel.

The MAGNABOND [9, 10] system makes the magnet physically and cryogenically stable under the considerable Lorentz force generated during operation. The energization of the magnet and adjustment of B is realized with a single controller (FUG NTS 1200M- 10 – Elektronik GmbH; including the power supply). In the event of a power failure, protective resistors, connected in series with diodes, limit the development of high voltages: if the breakdown voltage is exceeded, the protective circuit diverts a part of the current from the magnet windings and safely de-energizes the magnet.

### 2.1.2 Operation of the D-Mag

The magnet is energized and de-energized by a power supply. The maximum energization rate is 10 A/min. The maximum de-energization rate is 20 A/min. The magnet reaches its maximum field strength of 5.9 T after about 6 minutes and is de-energized after 3 minutes.

The thermometry for monitoring of the cool-down process and the actual temperature is done via Allen-Bradley resistors on the solenoid, which are read out by a multimeter. Two level meters (Intelligent Level Meter 211 - OXFORD Instrument) are to measure the remaining liquid content of the LN and LHe in their respective reservoirs. The sampling rate can be set from seconds (e.g. when filling up the cryostat) to several hours (during operation).

The manufacturer estimated the evaporation of helium during operation of the magnet to be about 0.25 l/h [10]. Our previous studies without running the cold head suggested an evaporation rate of about 0.76 l/h, necessitating daily refilling of the LHe cryostat. Additional losses of LHe due to the refilling process are strongly dependent on the handling of the operator and are estimated to be 7 - 10 l/refill by an experienced user. During the operation of the cold head, the evaporation decreases to 0.58 l/h (figure 6), whereby refilling of the LHe reservoir becomes necessary only every 36 hours. This extends the refilling intervals and thus further reduces the overall LHe consumption. Due to the geometry of the LHe reservoir (figure 3), it must be refilled before the level sinks below 70%. After the level has fallen below 70%, it takes another 3 hours until the critical temperature is exceeded and superconductivity is lost. A typical 250 liter LHe can is sufficient for a continuous operation of 9 – 11 days.

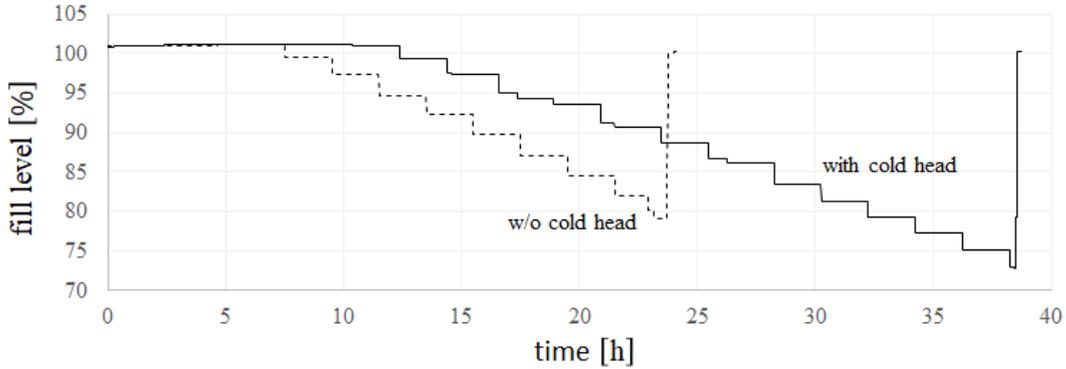

**Figure 6.** Comparison of LHe evaporation with and without operating cold head. With cold head the evaporation rate decreases from 0.76 L/h to 0.58 l/h. At a fill level of 79 %, respectively 73 % the LHe reservoir has been refilled. The level is checked every 30 minutes by the LHe probes to minimize the evaporation rate and stored by the DAQ software at 10 second intervals, resulting in the stepped pattern.

### 2.2 Analysis chamber

As indicated in the chapter 2.1.1, the gyrotron unit has been removed from the magnet vessel and replaced with a bakeable analysis chamber (figure 3). The chamber is 68.6 cm long and connected to the D-Mag vessel from above, via a custom welded flange, and sealed with a Viton® ring. A sample can be connected to a CF100 flange via the upper side of the analysis chamber, or it can be inserted through a CF63 flange via the lower side. The chamber is surrounded on the outside by a glass silk-sheathed copper heating cable, which is attached to the casing of the stainless steel tube by an adapted clamping rail. The chamber can be heated up via a bake-out unit. The melting point of the Viton® ring thermally limits the bake-out temperature of the analysis chamber to 150 °C. On the underside of the analysis chamber is a feed-through for the gas supply, via a welded-on Swagelok® connection, which can be used for direct venting, or for effective flushing of the chamber with specific gases.

### 2.3 Pumps and pressure control

The UHV-system presented here enables very precise pressure monitoring and pressure adjustments. The pumping station (Leybold Turbolab) of the D-Mag combines an air-cooled turbo-molecular-pump (TMP - Leybold Turbovac 350i) with a H$_2$ pumping speed of $S_{0,H}^{TMP}$ = 350 l/s and an oil-free scroll pump (SC - SCROLLVAC SC 15 D) with a maximum pumping speed of $S_{0,H}^{SC}$ = 15 m³/h. Two penning gauges are installed on the pumping station. One gauge (Leybold TTR 91 R), located between the SC and the TMP, serves as a pre-vacuum sensor for the TMP; a second high vacuum sensor (Leybold PTR 90 N) is attached to the cross-chamber, mounted atop of the TMP, and serves as a pressure control and safety instrument for the pump.

The pumping station is connected to the analysis chamber via several CF100 pipes with a total length of $l_{tot}$ = 276 cm from the upper flange of the TMP to the center of the solenoid. The gas is conducted through two cross chambers, which can be described as cylindrical elbow pipes, with $d$ = 10 cm, resulting in an extended length of additional 10 cm. This gives an effective length of $l_{eff.}$ = 286 cm. This results in a combined gas conductance of $C_{air}$ = 4.31 · 10⁻² m³/s of the evacuation pipe and an effective pumping speed of $S_{eff}$ = 38.4 l/s, according to:

$$S_{eff}^{-1} = S_0^{-1} + C_{air}^{-1} , \qquad (2.1)$$

where $S_0$ is the nominal pumping speed of the pump.

$$C_{air}^{-1} = \frac{12 \cdot l_{eff}}{\bar{c}_{air} \cdot \pi \cdot d^3} , \qquad (2.2)$$

with $d$ the diameter of the pipe segment and $\bar{c}_{air}$ is the mean thermal particle velocity of air at 273 K.

The entire pipe connection, including the cross chambers, is wrapped with a heating mat, which can be used to heat the UHV chambers to over 150 °C. After baking out the UHV-apparatus, the pump configuration allows to reach a residual gas pressure in the low 10⁻⁹ mbar range (N₂) within the analysis chamber. The purity and composition of the present vacuum is monitored by a residual gas analyzer (RGA - Pfeiffer Vacuum PRISMA pro QMA 250) in an atomic mass spectrum of up to 100 u and is equipped with a secondary electron amplifier.

A cold cathode ionization gauge (CC - MKS INST. 104220008) is located on top of the second cross chamber in front of the pumping station, and is used for readings in the range below 10⁻⁴ mbar. It is calibrated to partial pressures of He, N₂ and Ar, but can be manually recalibrated for any gas via the vacuum controller (946 Vacuum System Controller; type MKS 627f). Neutral gas pressure measurements are performed using two capacitance manometers/Baratrons® (CM - MKS Baratron Capacitance Manometer 627F 1MCD1B and 01MCC1B) with a combined measurement span ranging from 2 · 10⁻⁵ mbar up to 1 mbar (± 0,1%). One Baratron® is located together with the CC beyond the stray-field limit of 4mT, at the middle cross-chamber. The second CM is mounted directly above the magnet, so that a possible pressure gradient can be sampled via both neutral gas pressure measurements, which can be used to approximate the actual pressure within the analysis chamber empirically. This experimental setup is very valuable as the pressure difference in such a long UHV apparatus and in the molecular flow region can be significant.

During the operation of the magnet, the latter CM is also positioned at a safe distance from the magnetic field at the middle cross-chamber. The CMs are automatically zeroed with the CC at a pressure below 10⁻⁷ mbar, and the CC is in turn calibrated with the CMs in the jointly detectable pressure range (~10⁻⁴ mbar). In this way, the entire pressure range from 10⁻⁹ mbar to 1 mbar, including the transition range of the CC to the CM, can be acquired continuously and without jumps or breaks in the pressure readings.

A pneumatic vacuum slide valve (GV) is located between the center double cross-chamber and the cross chamber above the pumping station. A pressure gradient within the UHV apparatus can be circumvented by means of a 1 m long bypass, which diverts the residual gases past the GV via a 4 cm wide corrugated hose. The bypass significantly reduces the gas-conductance C = 7.7 l/s, but enables the pumps to be operated at 1 mbar without running hot, by reducing the gas flow rate.

One gas inlet is implemented directly at the bottom of the analysis chamber. Another inlet is located at the double cross chamber. Depending on the maximum pressure value and how sensitive a sample reacts to gas flows and pressure fluctuations, the user can choose between the two ports. The automatic regulation of the pressure is done by a freely adjustable PID-controlled Feedback loop on the vacuum controller. The set pressure is cross-checked with the current one via the CM readings, and regulated by an elastomer-sealed electrical flow controlled valve (MKS 0248D 00050RV), with a maximum flow rate of 50 sccm. The automatic pressure control functions in the pressure range of the two Baratrons® from $2 \cdot 10^{-5}$ mbar to 1 mbar and a corresponding software saves the measured values of the total pressure readout. Any pressure value below $2 \cdot 10^{-5}$ mbar is set manually and monitored with the CC. The CMs provide a digital data acquisition for of 3 Hz via a serial connection through the vacuum controller. For better temporal resolution, the voltage signals from the gauges can also be read out with an analog connection.

A Swagelok® tee joint is located upstream of the electrical valve to flush the gas supply lines, but can also be used to connect another gas pressure bottle to introduce gas mixtures into the analysis chamber.

## 2.4 Surface temperature measurement

The vacuum cross-chamber installed at the top of the D-Mag vessel is closed at the upper flange with a quartz-glass viewing window. A camera with visible and IR-sensitive sensors is mounted on a height-adjustable stand above the vessel and allows optical and thermal examination of the sample. For this purpose, we use two different cameras with adapted lenses. The cameras are mounted at a specific distance above the magnet so that their sensors are not disturbed by the magnetic stray field.

We use a high-resolution thermal camera (SensoTherm MV09), which operates in a spectral range between 0.75 µm and 1.08 µm, to examine structures below a size of 1 mm at a distance of 1 m from the measured object. This distance allows the infrared camera to be run during magnet operation. The detector of the camera has its highest resolution of 0.3 mm/pixel within 1100 °C to 3000 °C (± 1%) at a focal distance between 70 cm and 100 cm. Of course, this requires the knowledge of the emissivity of the material.

If the emissivity of the material under investigation is not known or uncertain, a dual-channel pyrometer can be used to obtain the true surface temperature. In special cases the correction factor has to be determined, because the emissivity changes between both wavelengths of the pyrometer (e.g. for lanthanum hexaboride). We use a fast (200 Hz) pyrometer (DIAS Pyrospot DSR 10NV) with a detection range of 800 - 2500 °C (± 0.5%). The dual-channel pyrometer measures in a range between 0.8 µm and 1.1 µm, with the short wavelength channel reaching its highest sensitivity at 0.96 µm and the long wavelength channel at 1.05 µm. The focal length of the lens allows the measurement of the temperature at a distance of up to 70 cm, with an average diameter of the object of 5 mm.

For a surface temperature reading by contacting, a 60 cm long sheath thermocouple on a CF100 flange with a corresponding feedthrough is available. The probe measures the temperature up to 1250 °C. In combination with the two-channel pyrometer or IR-camera, the correction factor of the sensor can be estimated easily, or, respectively, the emissivity of the material.

## 2.5 Residual gas analyzer

A residual gas analyzer (RGA – Pfeiffer Vacuum PRISMA pro QMA 250) is installed on the upper flange of the cross chamber above the pumping station. The RGA features a typical quadrupole mass spectrometer and a continuous secondary electron multiplier (c-SEM). The measuring range of the RGA is 1 u – 100 u and its minimum detection limit is $2.7 \cdot 10^{-15}$ mbar. In addition to residual gas analysis, the analyzer is also used to monitor the total pressure via its Faraday cup and the c-SEM. Detailed information on the RGA can be found in the vendor documentation [20].

## 3. Specific experiments in the D-Mag laboratory

Since the commissioning of the D-Mag laboratory in 2018 several experimental campaigns were conducted successfully. For example, experiments with dusty plasmas were carried out by A. Melzer et al. [1, 12]. Furthermore, investigations of Langmuir probes took place at the D-Mag in order to test the mechanical load under the influence of the magnetic field and thus the suitability for use in the Wendelstein 7-X fusion device. Experiments with novel neutral pressure gauges, equipped with crystal/ceramic and metal electron-emitters have also been performed on a regular basis in different gases. To provide some examples of the functionality and experimental capabilities of the D-Mag laboratory, we will present some of the results from the latter two topics mentioned in more detail. For the deeper physical understanding on Langmuir probes and crystal cathode pressure gauges (CCPG), the reader is referred to [19] and [8, 13], respectively.

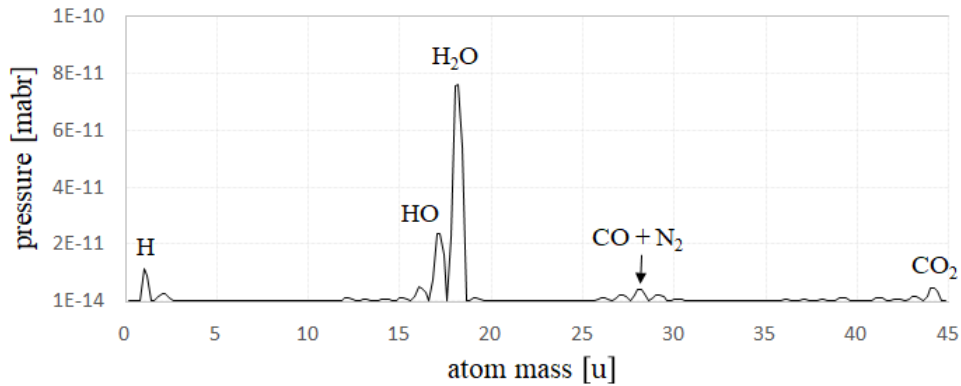

**Figure 7.** Characteristic residual gas spectrum after 3 hours of bakeout at 150 °C and 5 days of pumping. The total pressure is $2 \cdot 10^{-8}$ mbar; the oxygen partial pressure is negligible. Water is the main component of the residual gas molecules and can be removed further by prolonged bakeout.

### 3.1 Neutral pressure gauges with $LaB_6$ cathodes

A central motivation for setting up the D-mag laboratory was to provide a test stand for neutral pressure gauges. Neutral gas pressure gauges with a new cathode design, named crystal cathode pressure gauges (CCPG), which are based on the ASDEX pressure gauge (APG) [18], have been tested at the D-Mag laboratory in the magnetic field and in different gases. They are equipped with electron emitters made of lanthanum hexaboride ($LaB_6$). The CCPG has the advantage over conventional pressure gauges in that the cylindrical geometry of the cathode rod, and its low work function for electrons, functions even in strong magnetic fields and resists the usually disruptive Lorentz force. This makes it a key diagnostic for scrape-off-layer (SOL) physics in fusion devices, like W7-X: placed in front of the pumping ducts in the sub-divertor volume, the pressure measurement is a direct indicator of the exhaust rate [7].

The 8 mm long and 1 mm wide cylindrical emitter, made of $LaB_6$, is passively heated by contacting it with pyrolytic graphite plates (PGs). The PGs, unlike the $LaB_6$, have a high electrical resistance and are heated by ohmic heating. The electrodes, including a control electrode of the pressure gauge form a typical triode, which is arranged along the magnetic field lines (figure 9). The gauge has already been successfully used in operation phase 1.2 in W7-X in 2018 [7].

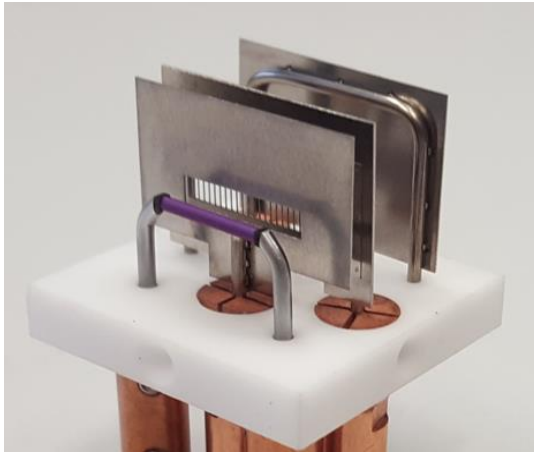
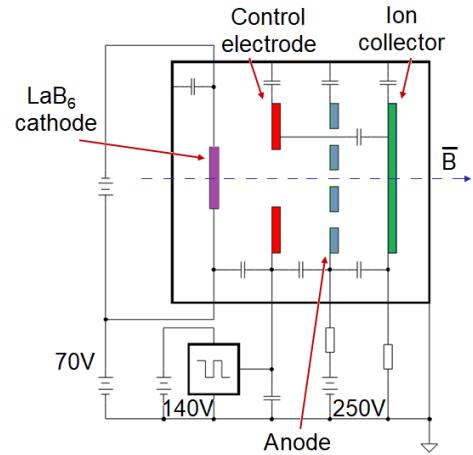

**Figure 8.** CCPG with a LaB6 (8 mm x 1 mm) cathode on an $Al_3O_2$ base plate with the triode configuration.

**Figure 9.** Schematic of the electronics with the potential settings: cathode (70 V), control electrode (140 V), anode (250 V), ion collector (0 V).

### 3.1.1 Thermal limit of operation

The following example is intended to demonstrate the camera's ability to perform precise thermal studies even with a strong magnetic field applied. Although the magnetic field itself should have no effect on the heat distribution on the emitter surface, we investigated the heat distribution within the emitter and PGs in more detail, using an IR camera Figure 10. The high resolution IR camera allows to study the temperature gradient on the emitter surface and also the temperature of the small PG blocks.

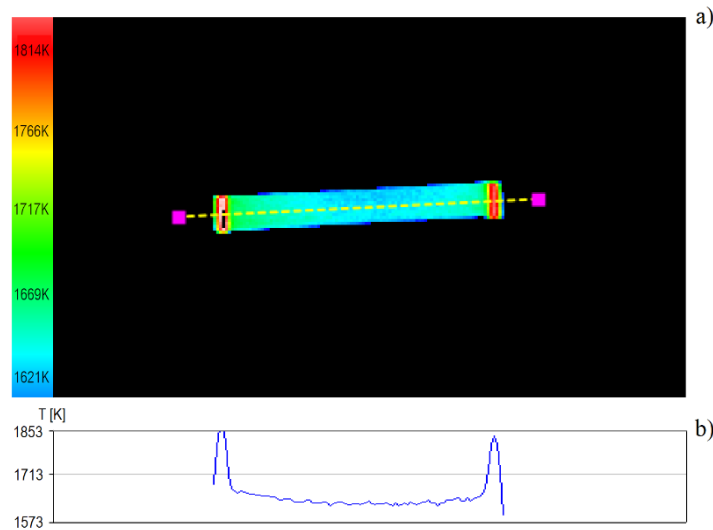

**Figure 10.** a) IR camera recording of the $LaB_6$ emitter in thermal equilibrium at a heating current of $I_{heat}$ = 2.4 A and in B = 3.1 T. b) Temperature profile following the dashed profile line or area of interest drawn in yellow. The thermal maxima are at the position of the PGs. A distinct thermal hollow profile has formed.

$LaB_6$ exhibits a relatively narrow temperature range for optimal operation. A too low temperature can lead to deposits on the emitter surface due to slow outgassing of impurities, slow diffusion and replenishment of LaB$_6$ to the emitter surface [14], which increases the work function, and ultimately to low electron emission. A too high temperature can negatively affect the lanthanum to boron ratio and can cause oxidation [15, 16] and melting of the emitter. Also the molybdenum-rhenium posts holding the cathode in place can bend (creeping) as a result of high thermal stress.

The wavelength-dependent emissivity of LaB$_6$ for the temperature regions relevant in this example has already been extensively investigated [17], so that relatively accurate measurement results may be assumed for this range, with a maximum error of 2%. The emitter exhibited a distinct hollow profile of the heat distribution. The temperature recording with the IR camera shown in figure 10 was measured at a heating current of $I_{heat}$ = 2.4 A at the emitter and at B = 3.1 T. Taking into account the transmissivity of the fused silica viewing window for IR wavelengths (~89%) and the emissivity of the pyrolytic graphite blocks ($\varepsilon$ = 0.80), a maximum temperature of 1780 °C was obtained at the PGs. An upper limit of 2.6 A for the heating current during prolonged operation of the gauges could therefore be determined. Above this limit, ductile alteration of the emitter rod itself and the Molybdenum-Rhenium mounts can occur during continuous operation.

### 3.1.2 Pressure steps and calibration

We operated and examined the CCPG in different gases, in a pressure range between $1 \cdot 10^{-7}$ mbar to 1 mbar. Calibration curves were recorded in hydrogen, deuterium and helium. The pressure gauge functioned properly up to a pressure range of 0.1 mbar. The precise pressure control made it possible to realize pressure steps and calibration curves (figure 11). These indicated, in the relevant pressure range from $2 \cdot 10^{-4}$ mbar to $1 \cdot 10^{-2}$ mbar, a quadratic dependence of the ion current on the pressure (P).

Sensitivity (S) is an important parameter for the characterization of pressure gauges and is defined as follows:

$$S = \frac{I_{ion}}{(I_{electron} - I_{ion}) \cdot P}, \qquad (3.1)$$

with the detected ion current $I_{ion}$, electron current $I_{electron}$, and the pressure $P$. The correction or subtraction of the ion current from the electron current in the denominator of equation 3.1 is relevant at high ion currents that occur at high pressures, especially with an applied magnetic field, which significantly increases the electron current density in the ionization volume. One of the reasons is the release of low-energy secondary electrons during gas ionization, which leads to an additional measured electron current without itself contributing to the ionization [18]. This leads to a nonlinear behavior of the gauges output over the neutral gas particle density, which can be corrected with the subtraction of the ion current in equation 3.1. It should be mentioned that the experiments here have been carried out in a pressure range in which the detected ion current is only about 1% of the electron current and therefore the correction plays only a minor role.

Sensitivity and temperature remained in an acceptable range. The determination of the sensitivity and temperature of the cathode, showed a steady decrease as the pressure increased. It is known that the sensitivity decreases with higher electron emission current [18].

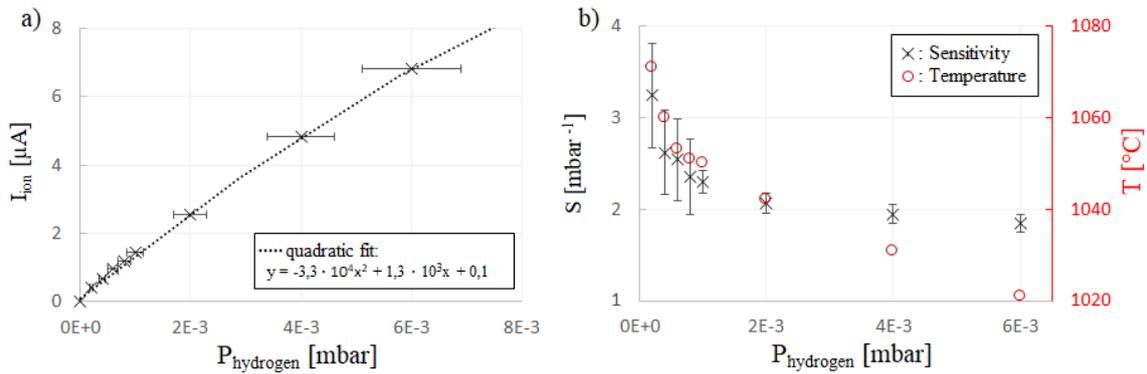

**Figure 11.** Results with an electron current of 625 µA and a magnetic field strength of 3 T. a) Ion current as a function of hydrogen pressure. b) Sensitivity of the pressure gauge and temperature of the LaB$_6$ cathode rod as a function of pressure.

### 3.2 Pressure calibration of the ITER pressure gauge prototype in magnetic field

The ITER fusion experiment requires pressure sensors that, once installed, must function for years in a magnetic field of up to 8 T. For this purpose, the ITER Diagnostics Team (ITED) in Garching, in collaboration with Fusion for Energy, has developed robust pressure gauges with an electron emitter, or cathode made of zirconium carbide [22].

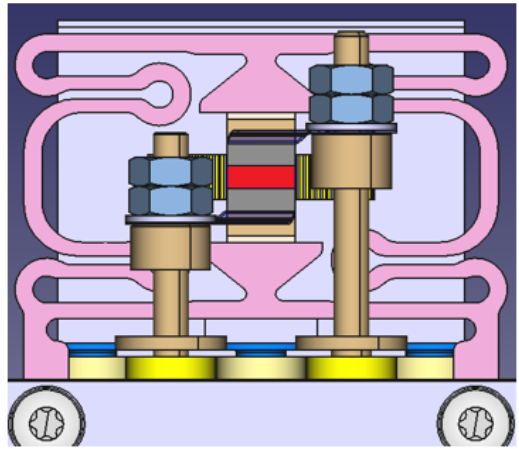

**Figure 12.** Left: CAD drawing of the pressure gauge head. Red – cathode (zirconium carbide), gray – pyrolytic graphite blocks, pink – holding frame, purple – control electrode.

The main difference to the CCPG, which is presented in chapter 3.1, is the attachment of the cathode made of zirconium carbide including the graphite blocks. For this purpose, following thermal model calculations and simulations, a ductile mounting clamp was constructed, which should also withstand very high thermal loads in the over long periods of time.

A prototype (figure 12) of the diagnostic pressure gauges (DPG) was tested in the D-Mag laboratory in early 2022, in residual gas, hydrogen and deuterium, and at a maximum magnetic field strength of 5.9 T.

We observed an increase of the ion current and the gauge's sensitivity, shown in figure 13, with increasing magnetic field strength: While without magnetic field the ion current was 0.51 µA, it jumped up to 0.72 µA after applying the coil current and increased to a maximum of 2.36 µA at 5.9 T during the progression of the magnetic field ramp. As a result, the sensitivity of the pressure gauge in the magnetic field increases by a factor of 4.6 at 5.9 T. According to established empirics, which we were able to confirm with $I_{electron}$ = 0.2 mA, the sensitivity should initially increase sharply up to a magnetic field strength of 1 T, but remain constant as the field strength increases further [18]. However, the plot of sensitivity versus magnetic field presented here, deviates from the foregoing description. This might be due to a different anode grid transparency, voltage distribution and electrode distances. We also have worked here with higher electron currents of 0.8 mA, so that present theoretical models for the case of $I_{electron}$ > 0.2 mA have to be examined more closely and, if necessary, modified.

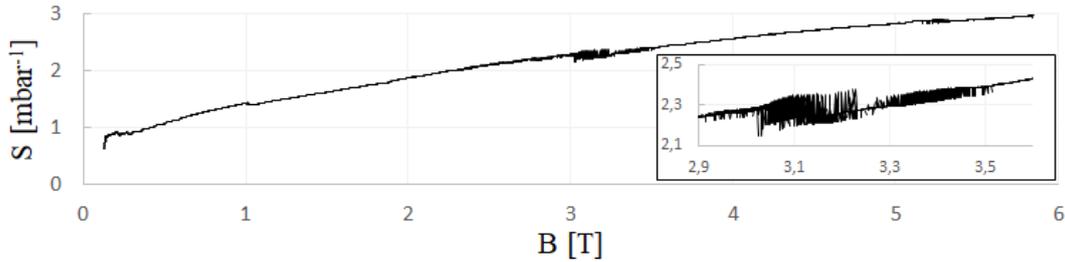

**Figure 13.** The sensitivity as a function of the magnetic field strength, in deuterium with fixed electron current of 800 µA and a constant pressure of $10^{-3}$ mbar. The sensitivity increases by a factor of 4.6 in the magnetic field. The crop-out shows signal bifurcations at 3.1 T.

The heating current increased from 3.90 A (without magnetic field) to 3.96 A, which is an acceptable operating point. Also noteworthy is the strong oscillations of the heating current at minutes 9 to 13. In this timeframe, at a field strength from 3 T to 3.5 T, jumps in the ion current appear ($\Delta I_{ion}$ = ± 6 % at 3.1 T). The exact cause of this ion current bifurcation has been investigated recently [21], where the author discusses several, possible underlying physical mechanism.

**3.3 Test of Langmuir probe driving assembly in the magnetic field**

The experimental campaign (OP 2) of the stellarator W7-X will have a new set of divertor Langmuir probes, to measure the plasma parameters at the divertor, such as the plasma density and electron temperature. Each probe is mechanically connected to a probe driving assembly at the back of the divertor for reciprocating the probes in and out of the plasma.

Before installing this reciprocating arrangement in W7-X, a prototype of the probe assembly along with the probe were tested in the D-Mag laboratory, which served as a test bed for this purpose. Figure 14 shows the layout of the prototype assembly in the resting position. The coil in the prototype installation was arranged vertically inside the analysis chamber of the D-Mag, so that the magnetic field vector points along the lengthwise orientation of the assembly geometry, to generate the necessary I × B force on the coil arms in the presence of a magnetic field. When the current through the coil is switched on, the coil experiences a torque, which makes the coil move. The coil can rotate about its axis, such that the maximum displacement of the probe tip can be 5 mm downward from its resting position, towards the plasma. A current flowing in the opposite direction would move the coil upwards, back to its resting position. A counterweight is attached to the assembly on the opposite side from the probe for testing the gravity assisted retraction.

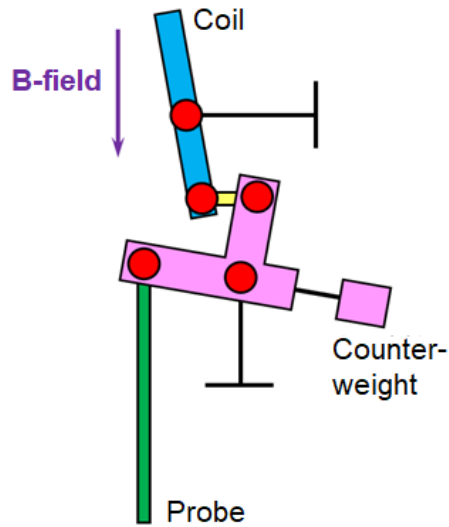

**Figure 14.** Simplified geometry of the coil assembly used in the D-Mag test for driving the divertor probe in and out of the plasma in W7-X. The coil carries a current, which leads to a torque on the coil moving it sideways. This motion of the coil makes the probe move up (retraction) and down (insertion), depending on the sign of the current flow.

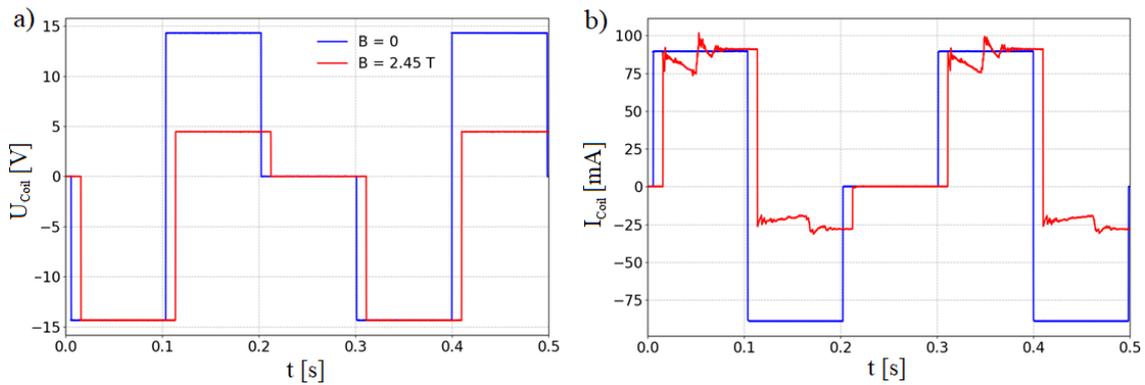

**Figure 15.** Voltage and current signals from two test runs in the D-Mag laboratory. a) Voltage signals for the cases with and without the magnetic field of the D-Mag. b) Current signals for the two cases.

The current through the coil and the voltage drop on the coil were recorded during the tests. The current flowing through the coil was changed by changing the input voltage to the coil. Due to the movement of the coil in the magnetic field an electromotive force (emf) was induced in the coil. This could be clearly seen in the current signal in presence of the magnetic field and was used to indicate the movement of the coil as well as to calculate the displacement of the probe tip. Voltage and current signals from two test runs are shown in figure 15. The plots in the figure on the left show the voltage drop on the coil for two test runs with and without magnetic field. The two runs had different input voltage and therefore different voltage drop on the coil. The plots on the right show the current flowing through the coil. It can clearly be seen from the right figure, that the case with magnetic field has fluctuations in the current signal whenever the voltage direction is reversed. These fluctuations correspond to the generation of an emf in the coil due to the movement of the coil. Once the movement stopped, the induced emf was zero. The induced emf signal was used to calculate the speed of the probe motion (angular velocity of the rotation of the coil) and subsequently to optimize the input voltage waveform such that the landing of the probe at the full insertion or full retraction positions was soft.

Many other tests were performed on the probe-drive assembly in the D-Mag laboratory, including tests with the heating mats of the analysis chamber switched on to simulate the operation in the heated condition of W7-X. The heating mats raised the temperature to ~150°C inside the chamber and hence the resistance of the probe-drive assembly. It was observed that a relatively larger input voltage was required to produce the same current and the same torque as in the absence of the heating.

## 4. Summary and conclusion

We have built a laboratory (D-Mag), which has a superconducting magnet and which is ideally suited for studies of plasma diagnostics in various gases, gas pressures, temperatures and high magnetic fields. Experiments can be performed in the pressure range between $10^{-9}$ mbar and atmospheric pressure. The magnetic field can be adjusted as desired and reaches a maximum field strength of 5.9 T. The temperature inside the analysis chamber can be raised in situ up to 150°C. After cooling and energizing the magnet, the helium consumption was 0.58 l/h when operating the cold head-compressor combination. Using a standard 250 liter LHe can, this allows an operating time of about 10 days until the next exchange of the can.

The D-mag laboratory allowed novel crystal cathode pressure gauges (CCPG) and a prototype ITER diagnostic pressure gauge (DPG) to be studied in more detail:
- Specific examples were used to demonstrate that pressure steps and temperature measurements can be performed to determine the sensitivity and thermal limits of the gauge. A maximum heating current of 2.6 A for long-term operation of the gauge, equipped with a $LaB_6$ emitter, was determined.
- Automated pressure ramps and uniform magnetic field ramps were realized. We found a nonlinear relationship between the sensitivity of the DPG and the magnetic field strength at electron currents above 0.2 A, which contradicts common theory; stating the sensitivity should stay stable with increasing magnetic field strength at B > 1 T [18]. The sensitivity increased steadily over the entire magnetic field ramp and reached a value 4.6 times higher at 5.9 T than without magnet operation.
- A prototype of a Langmuir probe drive assembly was successfully tested inside the D-Mag laboratory. It permitted the optimization of the input voltage waveform, such that the landing of the probe at the full insertion or full retraction positions is soft.

The described experiments allow the conclusion, that the D-Mag laboratory is a valuable tool for developing diagnostics in strong magnetic fields, as those encountered in fusion facilities. The laboratory setup is particularly suitable for pressure and magnetic calibrations of electrical instruments and detectors in a strong and variable magnetic field, such as the pressure gauges presented here.

## Acknowledgments

This work has been carried out within the framework of the EUROfusion Consortium, funded by the European Union via the Euratom Research and Training Programme (Grant Agreement No. 633053 and 101052200). The views and opinions expressed are however those of the author(s) only and do not necessarily reflect those of the European Union or the European

Commission. Neither the European Union nor the European Commission can be held responsible for them.

This work was partly supported by Fusion for Energy, in particular section 3.2, under the Grant F4E-FPA-364-SG06. The views expressed in this publication are the sole responsibility of the authors and do not necessarily reflect the views of Fusion for Energy and the ITER Organization. Neither Fusion for Energy nor any person acting on behalf of Fusion for Energy is responsible for the use, which might be made of the information in this publication.

Furthermore, we would like to thank A. Graband, for his active technical support during the assembly work and upgrading of the D-Mag laboratory, and M. Marquardt, for providing electro-technical assistance, programming and optimization of the laboratory soft- and hardware for the control and read-out of the CCPG.